\documentclass[envcountsame,
               runningheads]{llncs}

\usepackage{graphicx}
\usepackage{multirow}
\usepackage{latexsym}
\usepackage{amsmath,amssymb,amstext}
\usepackage{graphicx}
\usepackage{wasysym}
\usepackage{float}
\usepackage{enumerate}
\usepackage[arrow, matrix, curve]{xy}
\usepackage[justification=centering,singlelinecheck=off]{caption}
\usepackage{color}
\usepackage{url}
\usepackage{longtable}
\usepackage{wrapfig}
\usepackage{sidecap}
\usepackage{comment}
\usepackage{xcomment}
\usepackage{bussproofs}
\input{main.def}
\newcommand{\figpath}{figs}

\begin{document} 

\title{From Abstract Rewriting Systems\\ to Abstract Proof Systems}
   
\author{Clemens Grabmayer}
\institute{
    Universiteit Utrecht,
    Department of Philosophy\\
    Heidelberglaan 8,
    3584 CS Utrecht,
    The Netherlands\\
    \email{clemens@phil.uu.nl}
}
\maketitle


\begin{abstract}
  Some personal recollections on the introduction of `abstract
  proof systems' as a framework for formulating syntax-independent,
  general results about rule derivability and admissibility.
  With a particular eye on the inspiration I owe to Roel de Vrijer:
  the analogy with abstract rewriting systems.
  %
  %
  %
\end{abstract}


%
\begin{flushright}
  \emph{For Roel de~Vrijer with my best wishes\\
        on the occasion of his 60$^{\text{th}}$ birthday
       } 
\end{flushright}

\section{Introduction}
  \label{sec:intro}

As a first aim%
  \footnote{It turned out that work at this `first aim' eventually
            developed into my thesis \cite{grab:2005}.}
in my Ph.D.~project as \emph{AIO} at the VU-Amsterdam,
Jan Willem Klop suggested to me to investigate the 
proof-theoretic relationships between three kinds of proof systems for 
recursive type equivalence: 
a Hilbert-style proof system~\ACeq\ by Amadio and Cardelli (1993),
a coinductively motivated proof system~\BHeq\ by Brandt and Henglein (1998),
and a proof system~\AKeq\ by Ariola and Klop (1995)
for consistency-checking with respect to recursive type equivalence.
While originally formulated as a sequent-style calculus, the system of Brandt
and Henglein has a straightforward reformulation in natural-deduction style. 
The system of Ariola and Klop can be formulated as a tableau system. 

The easiest kind of relationship turned out to hold between the Brandt--Henglein 
and Ariola--Klop systems: 
proofs in \BHeq\ are, basically, mirror-images of 
`consistency unfoldings' (successful finite consistency-checks,
comparable to tableaux proofs) in \AKeq.%
  \footnote{%
    Jan Willem Klop had observed this mirroring property in examples
    before, and pointed me to this phenomenon.}
The proof-theoretic relationships of both systems with the Amadio--Cardelli system,
however, are considerably less straightforward. 
Although also in this case I found proof-transformations 
in both directions fairly soon, these transformations were quite complicated, 
and at least in part had the flavour of \emph{ad hoc} solutions.
The situation was theoretically unsatisfying insofar as a frequently
encountered situation, at the time of constructing these proof-transformations,
was the following.
While having shown that a particular rule~$\arule$ of a proof system~$\aprfsys_1$
can always be eliminated from a derivation in an extension~$\extd{\aprfsys_2}$
of another system~$\aprfsys_2$ by an effective procedure,
I was not sure at all whether applying such an elimination-procedure
was in fact essential. Perhaps instances of~$\arule$ could always
be modelled by derivations in~$\extd{\aprfsys_2}$,
and I only had failed to find some very easy demonstration of this fact.
In other words, by having only shown `admissibility' of~$\arule$ in~$\extd{\aprfsys_2}$,
I had not excluded the possibility that $\arule$ was actually `derivable'
in~$\extd{\aprfsys_2}$. 

For this reason I got interested in the relationship between, on the one hand,
the notions of rule derivability and rule admissibility,
and on the other hand, the precise manner of how rules can be eliminated
from derivations: either by `mimicking' derivations or by elimination procedures.
The notions of derivability and admissibility of inference rules date
back at least to 
Kleene \cite{klee:1952} (1952) and Lorenzen \cite{lore:1955} (1955),
and have since then been used for the analysis of concrete
proof calculi, and, this predominantly concerns rule admissibility, 
of what kinds of 
inferences a particular semantically given logic admits%
  \footnote{%
    An impressive body of work concerning this question is the study of
    admissible rules in intuitionistic propositional logic (IPC)
    by, to mention only a few names, Friedman, Rasiowa, Harrop, Rybakov, 
    Visser, de~Jongh, Ghilardi, and Iemhoff.%
  }.
Notwithstanding the familiarity of the concepts of rule derivability and admissibility,
the impression I got from the literature was that
these notions are usually defined only for concrete proof systems, and not 
in an abstract way. Furthermore, I did not find definitions that
applied to natural-deduction style systems directly (definitions which
I could immediately have applied for the Brandt--Henglein system \BHeq),
nor a presentation that explains the general practical relevance
of rule derivability and admissibility for \emph{interpretational proof theory},
and that is, for `syntactical translations of one formal theory into the other'
\cite{troe:schw:2000}.

However, I came across definitions of rule derivability and admissibility  
in the book \cite[p.70]{hind:seld:1986} of Hindley and Seldin, definitions that I found
appealing because they are based on an abstract notion of proof system
(`formal system'). 
While these definitions only pertain to Hilbert-style proof systems,
and while only basic properties are stated for rule
derivability and admissibility in \cite{hind:seld:1986},
two lemmas there (Lemma~6.14 and Lemma~6.15) stimulated me to think
about definitions for natural-deduction style systems,
and about how the mentioned lemmas could be extended.

In autumn 2002 I picked up again on my earlier triggered interest in
rule derivability and admissibility,
trying to investigate, in an abstract framework close to the one chosen
by Hindley and Seldin,
the relationship between 
derivability and admissibility of rules 
and 
the ways how rules can be eliminated from derivations.
Having gathered a number of, elementary, results about
derivability and admissibility in natural-deduction style proof systems,
I gave two talks on this topic in the regular TCS-seminar of our theory
group at the VU on Friday afternoons (January 24, and February 7, 2003).
While finding my results interesting, I recall that Roel de~Vrijer
reacted strongly against the concrete formulation of rules as extensional 
`rule descriptions' that I employed.

\paragraph{Overview.}
In Section~\ref{sec:rules} Roel's%
  \footnote{I hope to stay true to the spirit of this 
            \emph{liber amicorum} by switching to
            using only Roel de Vrijer's first name from now on.} 
remarks about two `naive' abstract notions of inference rule are explained,
which had highlighted to me significant problems of those notions. 
Section~\ref{sec:ARSs} is concerned with the concept
of `abstract rewriting system' in the theory of rewriting, which
Roel suggested me to look at.
In Section~\ref{sec:APHSs} the definition of `abstract pure Hilbert system'
(\APHS) is given and explained, which I formulated
as a direct consequence of Roel's remarks. 
In Section~\ref{sec:APHSs:results} derivability and admissibility of
rules in \APHS{s} are defined, and some results about these notions
are stated.  In Section~\ref{sec:ANDSs}
the idea of the notion of rules in `abstract natural-deduction systems'
is outlined. Finally in Section~\ref{sec:conclusion} a short summary
is given, and an idea for further research is mentioned.

\section{What is an inference rule, formally?}
  \label{sec:rules}

For this section heading I borrowed from the title
`\emph{What is an inference rule?}' of the article \cite{fagi:halp:vard:1992}
by Fagin, Halpern, and Vardi.
There, the investigation concentrates on the question of what
kinds of inferences semantically defined logics give rise to,
and how these inferences can be classified even with little specific knowledge
about the underlying semantics. 
Here, however, the focus is on the ways how rules are formulated
operationally, and how abstract rule definitions can be obtained.

Inference rules in logical calculi are defined in a variety of ways.
Here are just a few examples:
\begin{gather*}
   \begin{aligned}
      &
      \mbox{
         \AxiomC{$ A\to B\phantom{,} $}
         \AxiomC{$ A\phantom{,} $}
         \RightLabel{MP}
         \BinaryInfC{$ \rule{0pt}{2.2ex}
                        B $}
         \DisplayProof
            }
      & &
      \mbox{ 
         \AxiomC{$ [A]^{u} $}
         \noLine
         \UnaryInfC{$ \aderiv_1 $}
         \noLine
         \UnaryInfC{$ B $}
         \RightLabel{$\to$I, $u$}
         \UnaryInfC{$ A\to B $}
         \DisplayProof
           }
      & & 
      \mbox{
         \AxiomC{$ A[x \defdby t],\, \Gamma \Rightarrow \Delta $}
         \RightLabel{L$\exists$}
         \UnaryInfC{$ \rule{0pt}{2.2ex}
                      \exists x A, \, \Gamma \Rightarrow \Delta $}
         \DisplayProof
            }
   \end{aligned}
   \displaybreak[0]\\[1.5ex]
   \begin{aligned}
      &
      \mbox{
         \AxiomC{$ p_1 \overset{a}{\longrightarrow} p_2 $}
         \RightLabel{L$+$}
         \UnaryInfC{$ p_1 + q \overset{a}{\longrightarrow} p_2 $}
         \DisplayProof
            }
      & &   \hspace*{2ex}
      \mbox{
         \AxiomC{$ \tau_1 = \tau [ \alpha \defdby \tau_1 ] $}
         \AxiomC{$ \tau_2 = \tau [ \alpha \defdby \tau_2 ] $}
         \insertBetweenHyps{\hspace*{3ex}}
         \RightLabel{UFP}
         \BinaryInfC{$ \rule{0pt}{1.8ex}
                      \tau_1 = \tau_2 $}
         \DisplayProof
            }
   \end{aligned}
\end{gather*}
The rules in the first row are taken from proof calculi for,
in any case, classical predicate logic:
\emph{modus ponens} in a Hilbert-system, 
the $\to$-introduction rule in a natural-deduction system,
and the left-$\exists$-introduction rule in a sequent-style Gentzen system.
The rule {L$+$} in the second row is taken from a 
transition system specification~(TSS) of a process algebra
containing alternative composition~$+$,
and UFP is the unique-fixed-point rule 
in the Amadio--Cardelli system~\ACeq\ for recursive type equivalence.

As in the case of the examples above, rules in logical formalisms are
usually defined as schemes using a meta-language of the formula language
of the theory to be formalised such that the instances can be obtained
by meta-level substitution. 
In the easiest case of a Hilbert-style system (that is `pure', 
see Section~\ref{sec:APHSs}) a rule $\arule$ is usually represented by a
scheme expression of the form $A_1, \ldots, A_n / A$ that consists of 
premise expressions $A_1$, \ldots, $A_n$, and a conclusion expression
$A$, where $A_1, \ldots, A_n, A$ are expressions in an extension
of the formula language of the theory $\atheory$ to be formalised
with, for example, meta-variables for formulas and terms in $\atheory$,
and notation for object-language substitution in $\atheory$.  
The instances of $\arule$ are then defined to be the inferences 
of the form
$ (\funap{\Bar{\asubst}}{A_1})^*, \ldots, (\funap{\Bar{\asubst}}{A_n})^* /
  (\funap{\Bar{\asubst}}{\asubst})^* $
where $\asubst$ is a substitution 
that assigns formulas and terms of $\atheory$ 
to meta-variables of according sort,
$\Bar{\asubst}$ is a homomorphic extension of $\asubst$ to 
formula expressions including meta-variables,
and where $(\cdot)^*$ is an operation that evaluates formula expressions
containing substitution notation over the formula language of $\atheory$.
%

A precise formal introduction of an abstract notion of rule
that takes the syntactic view on rule definition seriously 
(along considerations just sketched)
seems to call for adopting concepts like the LF logical framework
\cite{harp:hons:plot}.
However, this was not the path on which I proceeded in work for
my thesis. Then I decided to leave the syntactic view on rule
definition out of consideration, and concentrate instead
on abstract notions of inference rule that do not require
syntactic restrictions to be imposed on the formula language. 

An abstract definition of rule is given by Hindley and Seldin
in \cite{hind:seld:1986}
in the form of `rule descriptions' 
(called `rules' in \cite{hind:seld:1986}):
partial functions that map sequences of premise formulas
to a single conclusion formula.

\begin{definition}\normalfont
  \label{def:ruledes:HS}
  Let $\forms$ be a set, and $n\in\nat$.
  A \emph{rule description} for an $n$-premise rule
  in a pure Hilbert system with set $\forms$ of formulas is a partial function
  $\saruledes \funin (\forms)^n \rightharpoonup \forms$.
  %
  %
  A rule description $\saruledes$ over $\forms$ 
  \emph{describes} the $n$-premise rule $\rulei{\saruledes}$ on $\forms$
  whose \emph{instances} are defined, for all $A_1,\ldots,A_n,A\in\forms$, by:
  \begin{align*}
    \mbox{
       \AxiomC{$ A_1 $}
       \AxiomC{$ \ldots $}
       \AxiomC{$ A_n $}
       \insertBetweenHyps{\hspace*{-0.75ex}}
       \TrinaryInfC{$ A $}
       \DisplayProof
          }
    \text{is an instance of $\rulei{\saruledes}$}
    \;\; & \Longleftrightarrow\;\;
    \aruledes{A_1,\ldots,A_n} = A \; .
  \end{align*}
\end{definition}

In the first of my earlier mentioned talks in 2003, 
I used rule descriptions according to this definition, 
and even generalisations for describing rules in
natural-deduction systems, to formulate results on rule derivability
and admissibility. 
During my presentation Roel immediately noticed, 
and acutely remarked, the obvious drawback of 
rule descriptions according to Definition~\ref{def:ruledes:HS}:
they are not able to describe rules that allow more than one conclusion 
to be inferred from a given sequence of premises. 
Rules such as for example:
\begin{align*}
  &
  \mbox{ 
     \AxiomC{$ A $}
     \RightLabel{$\vee$I$_R$}
     \UnaryInfC{$ \rule{0pt}{2ex} A \vee B $}
     \DisplayProof
        }
  & &
  \mbox{
     \AxiomC{$ \forall x A $}
     \RightLabel{$\forall$E}
     \UnaryInfC{$ \rule{0pt}{2ex} A [x \defdby t] $}
     \DisplayProof
        }
\end{align*}
the $\logor$-introduction rule, and the $\forall$-elimination rule
in natural-deduction systems for classical or intuitionistic logic.

Now it is an easy remedy to let rule descriptions for $n$-premise rules,
in the case of pure Hilbert systems, be defined as functions
$\saruledes \funin (\forms)^n \rightarrow \powerset{\forms}$,
and stipulate that such a rule description~$\saruledes$ 
defines the rule $\rulei{\saruledes}$  with the property
that $ A_1,\ldots,A_n / A $ is an instance of $\rulei{\saruledes}$
if and only if $A\in \aruledes{A_1,\ldots,A_n}$. 
While this is the change I carried out for my second talk,
it is perhaps more natural to drop the functional aspect in 
the formalisation of rule descriptions altogether,
and view rules as relations between premise formulas and a
conclusion formula.
Troelstra and Schwichtenberg have chosen such a formulation 
in \cite{troe:schw:2000} for the definition below of
rules in `LR-systems': systems
with `local rules' that correspond to pure Hilbert systems
(for the latter see the explanation at the start of Section~\ref{sec:APHSs}).

\begin{definition}
  \normalfont
  \label{def:rule:in:LRsystem}
  Let $\forms$ be a set, and $n\in\nat$.
  An $n$-premise \emph{LR-system rule} on (formulas of) $\forms$
  is a set of sequences $\seq{A_1,\ldots,A_n,A}$ in $\forms$ 
  of length $n+1$. Elements $\seq{A_1,\ldots,A_n,A}$ of
  such a rule are called \emph{instances} and usually written
  as $A_1,\ldots,A_n/A$. 
\end{definition}

While having carried out a change in the definition of rules
with the effect of Definition~\ref{def:rule:in:LRsystem},
I vividly remember that Roel had a further objection%
  \footnote{As far as I remember, this must have been in
    a discussion days after my second talk.}
to the changed definition: 
It is conceivable that a concrete syntactic definition of
a rule allows two or more instances with the same sequence of
premises and the same conclusion. There may be reasons
to preserve the existence of such extensionally,
but not intensionally, equivalent instances of rules. 

A possible example is the following formulation of an
$\logand$-elimination rule:
  %
  \begin{equation*}
     \mbox{ 
        \AxiomC{$ A_1 \wedge A_2 $}
        \RightLabel{$\wedge$E $(i\in\{1,2\})$}
        \UnaryInfC{$ \rule{0pt}{2ex} A_i $}
        \DisplayProof
          }
  \end{equation*}
for a proof system for propositional or predicate logic. 
This rule has two intensionally different instances:
  \begin{align*}
     &
     \mbox{ 
        \AxiomC{$ \bs{(x=0)} \wedge (x=0) $}
        \RightLabel{$\sAndElim$}
        \UnaryInfC{$ \rule{0pt}{2ex} \bs{x=0} $}
           \DisplayProof
           }
     & &
     \mbox{
        \AxiomC{$ (x=0) \wedge \bs{(x=0)} $}
        \RightLabel{$\sAndElim$}
        \UnaryInfC{$ \rule{0pt}{2ex} \bs{x=0} $}
        \DisplayProof
           }
  \end{align*}
whereas an LR-rule according to Definition~\ref{def:rule:in:LRsystem}
could only contain a single instance of the form
$\seq{x=0,x=0,x=0}$.

However, it has to be remarked that the rule $\sAndElim$ is usually split 
into the two parts $\sAndElimLeft$ and $\sAndElimRight$ 
that as their instances have the instances of $\sAndElim$
with $i$ chosen as $1$ or $2$, respectively.
Furthermore, it seems to me that logicians usually take care to avoid
rules that exhibit a behaviour similar as the $\logand$-elimination rule
above. At least I am presently unaware of a comparable example from 
`real-life' proof-theory. Being interested to hear from others, I formulate
the following question.

\begin{question}
  Do there exist generally accepted proof calculi that contain
  rules with instances that are extensionally, but not
  intensionally, equivalent?
\end{question}

I think that researchers in proof theory working on 
`deep inference' calculi, in which rules are allowed to
manipulate the syntactic structure of expressions not only
at their outer symbols but also inside of contexts
(therefore these proof systems possess a TRS-like feature),
could provide an answer quickly. 

Roel's second objection seemed to call for an entirely different framework
for abstract rule definitions. I took his suggestion to heart to
compare the situation with the definition of `abstract rewriting systems',
which had proved to be very useful in the theory of rewriting.

\section{Abstract rewriting systems}
  \label{sec:ARSs}

Rewriting theory knows many kinds of rewriting systems such as 
the $\lambda$-calculus, term rewriting systems (TRSs), combinatory reduction systems (CRSs),
higher-order rewriting systems (HRSs), graph rewriting systems,
and more.
In addition to the development of theory that is specific to 
each kind 
of rewriting system, it turned out to be very useful to develop also
an abstract rewriting theory, in which the concrete form or structure
of objects in a particular rewriting system is not taken into account. 
It turns out that even in a fully abstract framework
interesting statements can be formulated, for example, about
how properties of the reflexive-transitive closure $\sred^*$
of a rewrite relation $\sred$ follow from properties of $\sred$,
or from properties of the set of rewrite steps that induces $\sred$.  

Among the first concepts in abstract rewriting theory were the
results for `replacements systems' by Staples (1975),
and for `abstract reduction systems' by Klop (1980).
In their most simple formulation, abstract reduction systems
\cite{terese:2003} are defined as follows.

\begin{definition}
  \normalfont 
  \label{def:absredsys}
  An \emph{abstract reduction system} is a structure $\pair{A}{\sred}$
  consisting of a set $A$ with a binary \emph{reduction relation}~$\sred$
  on $A$.
\end{definition}

In a more general definition of `abstract rewriting system' 
(see also \cite{terese:2003}),
this notion denotes a system $\pair{A}{\{\sired{\alpha}\}_{\alpha\in I}}$ 
consisting of a set and a family of binary reduction relations on $A$. 
However, the results in \cite[Ch.~1]{terese:2003} bear witness 
to the fact that
even the simple version of abstract reduction system is a very fruitful
concept that leads to many interesting results, which due to their abstract
character can be applied to rewrite systems in general. 

But abstract rewriting systems in the simple form as
in Definition~\ref{def:absredsys}
have an evident drawback: 
they do not allow to distinguish between two different rewrite steps
from an object $a$ to an object $b$.

\begin{example}
  Consider the TRS $\atrs$ with as single rule
  %
     %
     $ \funap{f}{x} \red x $
     %
  %
  and with the induced rewrite relation $\ired{\atrs}$
  on the set ${\mit T\hspace*{-1pt}er}$ of terms
  over the signature $\Sigma = \{f\}$. 
  In this TRS, the term $\funap{f}{\funap{f}{a}}$ contains two redexes,
  which give rise to the steps:
  \begin{align*}
     \funap{\bs{f}}{\funap{f}{a}} &\red \funap{f}{a}
     &
     \funap{f}{\funap{\bs{f}}{a}} &\red \funap{f}{a} \; , 
  \end{align*}
  respectively.
  But in the extensional description of $\atrs$ 
  as an abstract reduction system $\pair{{\mit T\hspace*{-1pt}er}}{\sired{\atrs}}$,
  these steps cannot be distinguished,
  since both are witnessed by the same fact:
  $\funap{f}{\funap{f}{a}} \ired{\atrs}  \funap{f}{a}$.
  This phenomenon is called 
  a \emph{syntactic accident} (J.J. L\'{e}vy).
\end{example}

Preservation of the identity of rewrite steps under projecting to an 
abstract formalism is essential for proving abstract results on e.g.\
tracing steps under rewrite sequences, or on residuals of steps under other steps. 
This calls for an abstract framework in which `steps are first-class citizens'
(van~Oostrom), and where rewrite relations are only secondary notions. 
Such a framework is that of `abstract rewriting systems',
which date back to Newman \cite{newm:1942}, and in the formulation
by van~Oostrom and Roel de~Vrijer \cite{oost:vrij:2002,terese:2003}
are defined as follows.

\begin{samepage}
\begin{definition}
  \normalfont
  \label{def:absrewsys}
  An \emph{abstract rewriting system (ARS)} is 
  a quadruple of the form 
  $ \aARS = \quadruple{\aobjects}{\asteps}{\ssrc}{\stgt}$ 
  %
  where:
  %
  %
  \begin{itemize}
  \item[$\bullet$] $\aobjects$ and $\asteps$ are sets, 
     the set of \emph{objects}, and the set of \emph{steps}, of $\aARS$,
     respectively;
     %
     \vspace{0.75ex}
  \item[$\bullet$] 
    $ \ssrc, \stgt \funin \Phi \to A $ are functions,
    the {\em source\/} and the {\em target function\/} of~$\aARS$.
  \end{itemize}
  (See~Figure~\ref{fig:step:ARS} for a visualisation a step in an ARS
   as a hyperedge for a hypergraph.)
\end{definition}
\end{samepage}

\begin{Figure}[t]
  \begin{center}
    \scalebox{0.5}{\begin{picture}(0,0)%
\includegraphics{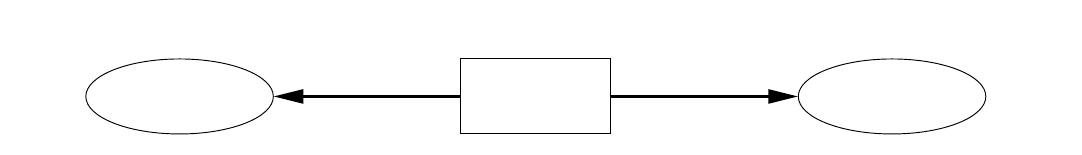}%
\end{picture}%
\setlength{\unitlength}{3947sp}%
\begingroup\makeatletter\ifx\SetFigFontNFSS\undefined%
\gdef\SetFigFontNFSS#1#2#3#4#5{%
  \reset@font\fontsize{#1}{#2pt}%
  \fontfamily{#3}\fontseries{#4}\fontshape{#5}%
  \selectfont}%
\fi\endgroup%
\begin{picture}(8649,1224)(1414,-5023)
\put(4351,-4411){\makebox(0,0)[b]{\smash{{\SetFigFontNFSS{17}{20.4}{\rmdefault}{\mddefault}{\updefault}{\color[rgb]{0,0,0}$\ssrc$}%
}}}}
\put(7051,-4411){\makebox(0,0)[b]{\smash{{\SetFigFontNFSS{17}{20.4}{\rmdefault}{\mddefault}{\updefault}{\color[rgb]{0,0,0}$\stgt$}%
}}}}
\put(5701,-4636){\makebox(0,0)[b]{\smash{{\SetFigFontNFSS{17}{20.4}{\rmdefault}{\mddefault}{\updefault}{\color[rgb]{0,0,0}$\astep$}%
}}}}
\put(8551,-4636){\makebox(0,0)[b]{\smash{{\SetFigFontNFSS{17}{20.4}{\rmdefault}{\mddefault}{\updefault}{\color[rgb]{0,0,0}$\tgt{\astep}$}%
}}}}
\put(2851,-4636){\makebox(0,0)[b]{\smash{{\SetFigFontNFSS{17}{20.4}{\rmdefault}{\mddefault}{\updefault}{\color[rgb]{0,0,0}$\src{\astep}$}%
}}}}
\end{picture}%
}
  \end{center}
  \caption{\label{fig:step:ARS}%
    Visualisation as a hypergraph hyperedge of
    a step~$\astep\in\asteps$ in an abstract rewriting system~%
    $ \aARS = \quadruple{\aobjects}{\asteps}{\ssrc}{\stgt}$
    with source formula $\src{\astep}\in\aobjects$
    and target formula $\tgt{\astep}\in\aobjects$.}
\end{Figure}

Newman \cite{newm:1942} attempted%
  \footnote{It is pointed out in \cite[cf.\ Rem.~6.14, (ii)]{oost:vrij:2002} 
            that Newman's attempt actually fails because for showing
            confluence of the $\lambda I$-calculus he uses
            statements that depend on abstract rewriting
            properties which do not hold for the $\lambda I$-calculus.} 
to show the confluence property of the
$\lambda I$-calculus by using abstract rewriting theory in systems
comparable to abstract rewriting systems.
An example of an important result that is based on abstract rewriting systems
is the theory of `abstract residual systems' \cite[Ch.~8]{terese:2003},
which allows to give uniform proofs for results 
in the theory of residuals
in orthogonal first-order and higher-order term rewriting systems (TRSs and HRSs).  

I think that it is likely that Roel's work as a co-author of
\cite{oost:vrij:2002} and his familiarity with abstract rewriting systems 
have informed his remarks to me about the problems with abstract definitions
of rules explained in the previous section.

\section{Abstract pure Hilbert systems}
  \label{sec:APHSs}

In this section the definition of `abstract pure Hilbert system' will be given,
which I developed starting in \cite{grab:2003} and then 
in my thesis \cite{grab:2005} as a reaction to and a direct consequence of
Roel's remarks as explained in Section~\ref{sec:rules}. 
Before turning to that definition, the concept of `pure Hilbert system'
needs to be explained.

In a strict interpretation of this concept, 
a `Hilbert system' is a proof system
for classical, intuitionistic, or minimal predicate (or propositional)
logic that consists of axiom schemes, and of the single rule
\emph{modus ponens}~(\ModusPonens):
\begin{center}
  \mbox{
    \AxiomC{$ \aform\to\bform $}
    \AxiomC{$ \aform $}
    \RightLabel{\ModusPonens}
    \BinaryInfC{$ \bform $}
    \DisplayProof
        }
\end{center}
In a more liberal interpretation, a Hilbert system may contain 
arbitrary rules~$\brule$ whose instances are of the form:
\begin{equation}
  \label{eq:pHS:inst}
  \mbox{
    \AxiomC{$ \aform_1 $}
    \insertBetweenHyps{\;\;\ldots\;\;}
    \AxiomC{$ \aform_n $}
    \RightLabel{$\brule$}
    \BinaryInfC{$ \aform $}
    \DisplayProof
        }
\end{equation}

A Hilbert system~$\aHS$ (in the liberal sense)
is called `pure' if
rule applications within derivations in $\aHS$ do not depend
on the presence or absence of assumption occurrences in
the immediate subderivations. More precisely, the following
must hold:
Suppose that $\brule$ is a rule in a Hilbert system $\aHS$.
And suppose that, for some formulas
$ \aform, \aform_1, \ldots, \aform_n $ of $\aHS$, 
\eqref{eq:pHS:inst} is an instance of $\brule$. Furthermore, 
let $ \aderiv_1, \ldots, \aderiv_n $
be derivations in $\aHS$ (which may contain unproven assumptions) 
with conclusions $ \aform_1, \ldots, \aform_n $, respectively.
Then the prooftree $\aderiv$ of the form:
\begin{center}
  \mbox{
    \AxiomC{$ \aderiv_1 $}
    \noLine
    \UnaryInfC{$ \aform_1 $}
    \AxiomC{$ \aderiv_n $}
    \noLine
    \UnaryInfC{$ \aform_n $}
    \insertBetweenHyps{\;\;\ldots\;\;}
    \RightLabel{$\brule$}
    \BinaryInfC{$ A $}
    \DisplayProof
       }
\end{center}
is a derivation in $\aAPHS$, irrespectively of whether
or not assumptions occur in $ \aderiv_1, \ldots, \aderiv_n $,
and how these do actually look like. 

Typical examples of pure Hilbert systems are Hilbert systems
in the strict sense for propositional or predicate logic. 
Examples of non-pure Hilbert systems are the well-known
proof systems for modal logics such as 
\modlogicK, \modlogicT, \modlogicSfour, and \modlogicSfive,
which contain the \emph{necessitation rule}~(\Necessitation)
with instances of the form:
\begin{equation}
  \label{eq:NR}
  \mbox{
    \AxiomC{$ \aform $}
    \RightLabel{\Necessitation}
    \UnaryInfC{$ \necessite{\aform} $}
    \DisplayProof
       }
\end{equation}
The applicability of \Necessitation\
in derivations of these proof systems is restricted to situations
in which the immediate subderivation of the premise 
does not contain any unproven assumptions 
(and hence this premise must be a theorem).
Therefore instances of \Necessitation~of the form \eqref{eq:NR} 
are subject to a side-condition on the absence
of assumptions in the immediate subderivation of the premise,  
and hence Hilbert systems including this rule are impure, that is, not pure.

In the stipulation for the property `pure' of Hilbert systems I followed
an analogous stipulation that Avron uses in \cite{avro:1991} for 
`Hilbert-type systems for consequence'.
There, a nice characterisation of Hilbert systems and
natural-deduction systems among all sequent-style Gentzen systems is given.
In Avron's terminology, 
pure Hilbert systems as explained above correspond%
  \footnote{This correspondence is made explicit and proved in
            Appendix~E of \cite{grab:2003}.}
to pure, single-conclusioned 
Hilbert-type systems for consequence, which formally are sequent-style systems.

Pure Hilbert systems are among the easiest kinds of proof systems.
Therefore it was natural for me to start with formulating 
an abstract version of these systems, before later going on to define 
abstract versions of systems of natural deduction. 
The concept of `abstract pure Hilbert system (\APHS)', 
which will be defined below,
is analogous to abstract rewriting systems in the sense that steps
between objects (here, inference steps between formulas) are central, not
the relations between objects (but these relations are induced by the steps).
However, somewhat different from the definition of ARSs,
in which the notion of a (rewrite) rule does not figure at
all, in the concept of an \APHS\ the distinction between different
inference rules is retained. The foremost reason for this was 
my wish to let derivations in \APHS{s} closely resemble
prooftrees in a usual sense with rule names displayed next
to the derivations' inferences. 

As a consequence it is a prerequisite for the definition of \APHS{s} 
(see Definition~\ref{def:APHS} below) to introduce first the 
concept of `unnamed \APHS-rule'.
Here, the following notation will be used: 
For all sets $X$, by $\finseqs{X}$ the set of all finite-length sequences
of elements of $X$ is meant; 
the empty sequence is denoted by $\seq{}$, and $n$-element sequences 
in $\finseqs{X}$ are written as $\seq{x_1,\ldots,x_n}$,
where $x_1,\ldots,x_n\in X$.

\begin{definition}\normalfont 
  \label{def:APHS:rule}
  Let $\forms$ be a set.
  An {\emph{unnamed \APHS-rule}}
     (an unnamed rule for an `abstract pure Hilbert system')
  on (the formulas of) $\forms$ is a triple 
  $ \aunnamedRule = \triple{\insts}{\sprem}{\sconcl}$ 
  where:
  %
  %
  \begin{itemize}
    \item[$\bullet$]  
      $\insts$ is a set whose elements are called
      the {\em instances\/} of $\aunnamedRule$,
      \vspace{0.75ex}
    \item[$\bullet$] 
      $\sprem \funin \insts \to \finseqs{\forms}$ and
      $\sconcl \funin \insts \to \forms$
      are the {\em premise function\/}
      and the {\em conclusion function\/} of $\aunnamedRule$, respectively.
  \end{itemize}
  For all sets $\forms$, by $\APHSrules{\forms}$ 
  the class of unnamed \APHS-rules on $\forms$ is denoted.
\end{definition}

Note that unnamed \APHS-rules are allowed to have instances with different
arities. 
In addition to the functions $\sprem$ and $\sconcl$ 
associated with a rule,
for an unnamed \APHS-rule $\aunnamedRule = \triple{\insts}{\sprem}{\sconcl}$
\begin{align*}
   \text{the function} \quad
      & \sarity : \insts \rightarrow \,\nat \;\,
      & & \text{and}
                                                                 \\[-0.75ex]
   \text{the partial functions} \quad
      & \spremi{i} : \insts \rightharpoonup \forms 
      & & \hspace*{-6ex}
        \text{(for all $i\in \nat $)}  
\end{align*}
are defined as follows:
$\sarity$ assigns to an instance $\ainst$ of $\aunnamedRule$ the number of
its premises, that is, the length of the sequence $\prem{\ainst}$.
For all $i\in\nat, i\ge 1$,
$\premi{i}{\ainst}$ assigns to an instance $\ainst$ of $\aunnamedRule$
its $i$-th premise, that is, the $i$-th formula $\aform_i$ in
$ \prem{\ainst} = \seq{\aform_1,\ldots,\aform_i,\ldots} $ if it exists;
otherwise, $\premi{i}{\ainst}$ is undefined. 
\begin{Figure}[t]
  \begin{center}
    \scalebox{0.5}{\begin{picture}(0,0)%
\includegraphics{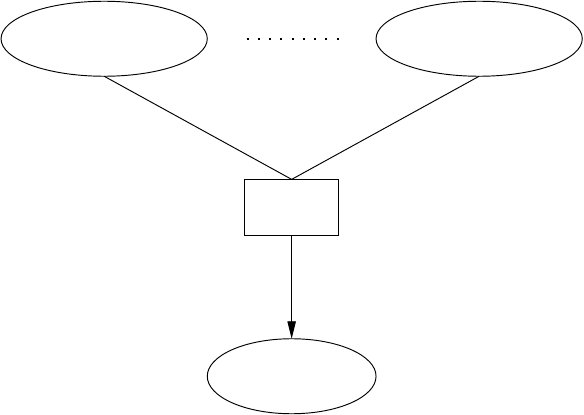}%
\end{picture}%
\setlength{\unitlength}{3947sp}%
\begingroup\makeatletter\ifx\SetFigFontNFSS\undefined%
\gdef\SetFigFontNFSS#1#2#3#4#5{%
  \reset@font\fontsize{#1}{#2pt}%
  \fontfamily{#3}\fontseries{#4}\fontshape{#5}%
  \selectfont}%
\fi\endgroup%
\begin{picture}(4666,3314)(2618,-5393)
\put(3451,-2461){\makebox(0,0)[b]{\smash{{\SetFigFontNFSS{17}{20.4}{\rmdefault}{\mddefault}{\updefault}{\color[rgb]{0,0,0}$\premi{1}{\ainst}$}%
}}}}
\put(6451,-2461){\makebox(0,0)[b]{\smash{{\SetFigFontNFSS{17}{20.4}{\rmdefault}{\mddefault}{\updefault}{\color[rgb]{0,0,0}$\premi{n}{\ainst}$}%
}}}}
\put(4951,-5161){\makebox(0,0)[b]{\smash{{\SetFigFontNFSS{17}{20.4}{\rmdefault}{\mddefault}{\updefault}{\color[rgb]{0,0,0}$\concl{\ainst}$}%
}}}}
\put(4276,-3061){\makebox(0,0)[b]{\smash{{\SetFigFontNFSS{14}{16.8}{\rmdefault}{\mddefault}{\updefault}{\color[rgb]{0,0,0}$1$}%
}}}}
\put(5626,-3061){\makebox(0,0)[b]{\smash{{\SetFigFontNFSS{14}{16.8}{\rmdefault}{\mddefault}{\updefault}{\color[rgb]{0,0,0}$n$}%
}}}}
\put(4951,-3811){\makebox(0,0)[b]{\smash{{\SetFigFontNFSS{17}{20.4}{\rmdefault}{\mddefault}{\updefault}{\color[rgb]{0,0,0}$\ainst$}%
}}}}
\end{picture}%
}
  \end{center}
  \caption{\label{fig:inst:APHS-Rule}%
           Visualisation as hypergraph hyperedge of
           an instance $\ainst$ of an \APHS-rule
           with arity $\arity{\ainst} = n$.}
\end{Figure}

Using these definitions, a visualisation 
of an instance $\ainst\in\insts$ with $\arity{\ainst} = n$
of an unnamed \APHS-rule 
$\aunnamedRule = \triple{\insts}{\sprem}{\sconcl}$
as a hypergraph hyperedge
is given in Figure~\ref{fig:inst:APHS-Rule}.
In a prooftree such an instance can be written as: 
\begin{center}
  \mbox{
    \AxiomC{$ \premi{1}{\ainst} $}
    \insertBetweenHyps{\;\;\ldots\;\;}
    \AxiomC{$ \premi{n}{\ainst} $}
    \BinaryInfC{$ \concl{\ainst} $}
    \DisplayProof
        }
\end{center}

While rules \emph{for} \APHS{s} do not carry names,
rules \emph{in} abstract pure Hilbert systems as defined below
\emph{do} carry names.

\begin{definition} \normalfont
   \label{def:APHS}
  An \emph{abstract pure Hilbert system (an \APHS)~$\aAPHS$} is a quadruple
  $\quadruple{\forms}{\names}{\namedAxioms}{\namedRules}$ 
  where:
  \begin{itemize}
    \item[$\bullet$]  
      $\forms$, $\names$, $\namedAxioms$ and $\namedRules$ are sets
      whose elements are called
      the {\em formulas of\/} $\aAPHS$,
      the {\em names\/} (for axioms and rules) {\em in\/} $\aAPHS$,
      the {\em named axioms of\/} $\aAPHS$,
      and the {\em named rules of $\aAPHS$\/}, respectively;
      we demand $ \forms \neq \emptyset $, i.e.\ that
      the formula set be nonempty;
    \item[$\bullet$] 
      $ \namedAxioms \,\subseteq\, \forms \times \names$, 
      i.e.\ the named axioms of $\aAPHS$ are tuples with
      formulas of $\aAPHS$ as first, and names in $\aAPHS$
      as second, components;
    \item[$\bullet$] 
      $\namedRules \,\subseteq\, \APHSrules{\forms} \times \names$,
      i.e.\ the named rules of $\aAPHS$ are tuples that have
      unnamed \APHS-rules on $\forms$ as their first, and names
      as their second components; 
    \item[$\bullet$] 
      for the named axioms and the named rules of $\aAPHS$
      the following holds:
      \begin{enumerate}
        \renewcommand{\labelenumi}{(\roman{enumi})}
        \item  names of named axioms in $\namedAxioms$ are different from names 
          of named rules,
        \item different named rules in $\namedRules$ carry different names
          (but for an unnamed APHS-rule $\aunnamedRule$ and different
           names $\aname_1, \aname_2 \in\names$, it is possible that    
           $\pair{\aunnamedRule}{\aname_1}, \pair{\aunnamedRule}{\aname_2}\in\namedRules$).
          %
      \end{enumerate}
      %
     %
   \end{itemize}    
\end{definition}

This definition of \APHS{s}, which is essentially the formulation in
my thesis \cite{grab:2005}, is actually a slight reformulation of the notion of
`abstract Hilbert system with names for axioms and rules' (\nAHS)
in \cite{grab:2003} after discussions with Roel in my last year
as Ph.D.~student, when I was writing down my thesis.
In \cite{grab:2003} an \nAHS\ is defined as 
a quintuple $\quintuple{\forms}{\axioms}{\rules}{\names}{\sname}$
consisting of a set $\forms$ of formulas, a set $\axioms\subseteq\forms$
of axioms, a set $\rules\subseteq\APHSrules{\forms}$ of unnamed \APHS-rules,
a set $\names$ of names, and a name function
$\sname \funin \rules \to \names$.
Roel remarked that this use of a name function on a set of unnamed \APHS-rules 
prevents the possibility that the same unnamed \APHS-rule could occur twice,
under different names, in an \APHS. He convinced me that the formalisation should
not a priori exclude systems in which an initially unnamed rule occurs
under more than one name. 

Let
$ \aAPHS = \quadruple{\forms}{\names}{\namedAxioms}{\namedRules}$
be an APHS.
In prooftrees, an instance $\ainst\in\insts_{\aunnamedRule}$ of 
an unnamed rule $\aunnamedRule$ such that 
$\pair{\aunnamedRule}{\aname} \in \namedRules$, 
is written as:
\begin{center}
  \mbox{
    \AxiomC{$ \premi{1}{\ainst} $}
    \insertBetweenHyps{\;\;\ldots\;\;}
    \AxiomC{$ \premi{n}{\ainst} $}
    \RightLabel{$\aname$}
    \BinaryInfC{$ \concl{\ainst} $}
    \DisplayProof
        }
\end{center}
A \emph{derivation in $\aAPHS$} is defined as a prooftree in
the sense of \cite{troe:schw:2000}:
a tree in which the nodes are labelled by formulas and in which
the edges make part of rule instances and are not drawn,
but are replaced by horizontal lines that represent rule instances
and carry the name of the rule that is applied. 
Axioms and assumptions appear as top
nodes; lower nodes are formed by applications of rules;
the bottommost formula is the conclusion.
If $\aderiv$ is a derivation in $\aAPHS$, then 
by $\assm{\aderiv}$ the set of assumptions of $\aderiv$,
and by $\concl{\aderiv}$ the conclusion of $\aderiv$ will be denoted. 
Let 
$\aunnamedRule$ be an \APHS-rule on the formula set of $\aAPHS$. 
A derivation $\aderiv$ in $\aAPHS$ is called 
a \emph{mimicking derivation} for an instance $\ainst$ of $\aunnamedRule$ 
if
$\assm{\aderiv}\subseteq\set{\prem{\ainst}}$ 
and $\concl{\aderiv} = \concl{\ainst}$.

The (usual, standard) consequence relation $\sconsrelH{\aAPHS}$
in an \APHS~$\aAPHS$ with formula set $\forms$ 
is a binary relation between sets of formulas and formulas
of $\aAPHS$.
For all formulas $\aform\in\forms$, and sets $\asetforms\subseteq\forms$
of formulas, in $\aAPHS$,
$\consrelH{\aAPHS}{\asetforms}{\aform}$ holds
if there is a derivation $\aderiv$ in $\aAPHS$ such that
the assumptions of $\aderiv$ are contained in $\asetforms$,
and $\aform$ is the conclusion of $\aderiv$.
By $\consrelH{\aAPHS}{}{\aform}$ 
the statement $\consrelH{\aAPHS}{\emptyset}{\aform}$ is abbreviated.
A formula $\aform$ is a \emph{theorem} of an \APHS~$\aAPHS$ if
$\consrelH{\aAPHS}{}{\aform}$ holds, that is, if there is a derivation
in $\aAPHS$ without assumptions that has $\aform$ as its conclusion.

Let $\aAPHS$ be an \APHS,
$\aunnamedRule$ an unnamed \APHS-rule over the set of formulas of $\aAPHS$,
and $\anamedRule = \pair{\aunnamedRule}{\aname}$  
a named version of $\aunnamedRule$.
Then the result of adding $\anamedRule$ to the named rules of $\aAPHS$ is
denoted by $\extsysbyrule{\aAPHS}{\anamedRule}$ given that it is an \APHS.

\section{Rule derivability and admissibility in \APHS{s}}
  \label{sec:APHSs:results}

In this section the definitions of derivability and admissibility
of arbitrary \APHS-rules in an \APHS\ are introduced,
based on the stipulations at the end of the previous section.
Furthermore, basic results about these notions 
are stated, together with results that I have obtained
in \cite{grab:2003} and \cite{grab:2005}.

In the definitions below of rule derivability and admissibility in \APHS{s}
the following notation will be used:
for a set $X$ and a sequence $\aseq\in\finseqs{X}$, $\set{\aseq}$
denotes the set of elements of $X$ that occur in $\aseq$.

Let $\aAPHS$ be an abstract Hilbert system,
and let
$ \aunnamedRule = \triple{\insts}{\sprem}{\sconcl}$
an unnamed \APHS-rule over the set of formulas of $\aAPHS$.
Then $\aunnamedRule$ is called \emph{derivable in $\aAPHS$} if:
\begin{equation}
  \text{for all $\ainst\in\insts$:} \;\;
    \consrelH{\aAPHS}{\set{\prem{\ainst}}}{\concl{\ainst}} \; .
\end{equation}
$\aunnamedRule$ is called \emph{correct for $\aAPHS$} if: 
\begin{equation}
  \text{for all $\ainst\in\insts$:} \;
    \bigl[\,
    (\text{for all $\aform\in\set{\prem{\ainst}}$: 
     $\consrelH{\aAPHS}{\,}{\aform}$})
       \;\Longrightarrow\;
         \consrelH{\aAPHS}{}{\concl{\ainst}} 
    \,\bigr] \; .
\end{equation}
And $\aunnamedRule$ is stipulated to be \emph{admissible in $\aAPHS$} if
every (or equivalent: one)
extension $\extsysbyrule{\aAPHS}{\anamedRule}$ of $\aAPHS$
by adding a named version $\anamedRule$ of $\aunnamedRule$
has the same theorems as $\aAPHS$. 
A named rule $\anamedRule = \pair{\aunnamedRule}{\aname}$ 
is derivable/correct/admissible in $\aAPHS$
if the underlying unnamed rule $\aunnamedRule$ 
is derivable/correct/admissible in $\aAPHS$, respectively.

Each of these definitions can be reformulated in terms of `mimicking derivations'.
For example,
an unnamed \APHS-rule $\aunnamedRule$ is derivable in an \APHS~$\aAPHS$ 
if and only if
for every instance of $\aunnamedRule$ there exists a mimicking derivation
in $\aAPHS$.

The following proposition gathers some of the most basic properties of
rule derivability, correctness, and admissibility, and of their interrelations.
Items~(i)--(iii) of this statement are a reformulation for \APHS{s} of 
a lemma by Hindley and Seldin (Lemma~6.14 on p.~70 in \cite{hind:seld:1986}).
Item~(iv) is taken from \cite{grab:2003} (see Theorem~3.5 on p.~18--19 there).

\begin{proposition}
  Let $\aAPHS$ be an \APHS, and let $\aunnamedRule$ be an unnamed \APHS-rule
  on the set of formulas of $\aAPHS$. Then the following statements hold:
  \begin{enumerate}
    \renewcommand{\labelenumi}{(\roman{enumi})}
    \item $\aunnamedRule$ is admissible in $\aAPHS$ if and only
      if $\aunnamedRule$ is correct for $\aAPHS$.
    \item If $\aunnamedRule$ is derivable in $\aAPHS$, 
      then $\aunnamedRule$ is admissible in $\aAPHS$. But the converse
      implication does not hold in general.
    \item If $\aunnamedRule$ is derivable in $\aAPHS$, the
      $\aunnamedRule$ is derivable in every extension of $\aAPHS$ by
      adding new formulas, new axioms, and/or new rules. 
    \item $\aunnamedRule$ is derivable in $\aAPHS$ if and only if
      $\aunnamedRule$ is admissible in every extension of $\aAPHS$ by
      adding new formulas, new axioms, and/or new rules. 
  \end{enumerate}
\end{proposition}

Further results that I have obtained in \cite{grab:2003} include the
following:
\begin{enumerate}
  \item
    Let $\aAPHS$ be an \APHS, $\aunnamedRule$ an unnamed \APHS-rule,
    and $\extsysbyrule{\aAPHS}{\anamedRule}$ an extension of $\aAPHS$
    by adding a named version $\anamedRule = \pair{\aunnamedRule}{\aname}$
    of $\aunnamedRule$. 
    If $\aunnamedRule$ is admissible in $\aAPHS$,
    then every derivation $\aderiv$ in $\extsysbyrule{\aAPHS}{\anamedRule}$ 
    without assumptions can
    be replaced by a mimicking derivation $\aderiv'$
    in $\aAPHS$ (without assumptions).
    If $\aunnamedRule$ is derivable in $\aAPHS$,
    then every derivation $\aderiv$ in $\extsysbyrule{\aAPHS}{\anamedRule}$ can
    be replaced by a mimicking derivation $\aderiv'$ in $\aAPHS$, which
    moreover can be found by stepwise replacements 
    of $\anamedRule$-instances
    in $\aderiv$ by mimicking derivations in $\aAPHS$. 
  \item
    Let $\aAPHS_1,\aAPHS_2$ be \APHS{s} that have the same set of formulas.
    The following statements are equivalent with 
    the statement that
    $\aAPHS_1$ and $\aAPHS_2$ have the same admissible rules:
    (i)~the rules of $\aAPHS_1$ are admissible in $\aAPHS_2$,
    and vice versa,
    and
    (ii)~$\aAPHS_1$ and $\aAPHS_2$ have the same theorems.
    Statements equivalent with the assertion that
    $\aAPHS_1$ and $\aAPHS_2$ have the same derivable rules
    are:
    (i)$'$~the rules of $\aAPHS_1$ are derivable in $\aAPHS_2$,
    and vice versa,
    and
    (ii)$'$~$\aAPHS_1$ and $\aAPHS_2$ induce the same consequence
    relation.
  \item
    Let $\aAPHS$ be an \APHS, $\aunnamedRule$ an unnamed \APHS-rule,
    and $\extsysbyrule{\aAPHS}{\anamedRule}$ an extension of $\aAPHS$
    by adding a named version $\anamedRule = \pair{\aunnamedRule}{\aname}$
    of $\aunnamedRule$. 
    If $\aunnamedRule$ is derivable in $\aAPHS$,
    then $\anamedRule$-elimination for derivations in 
    $\extsysbyrule{\aAPHS}{\anamedRule}$ can be performed effectively: 
    For a given derivation $\aderiv$ in $\extsysbyrule{\aAPHS}{\anamedRule}$,
    pick an arbitrary instance of $\anamedRule$ in the derivation
    and replace it, in the derivation,  by a mimicking derivation;
    carry out such {\em mimicking steps\/} repeatedly until no
    further applications of $\anamedRule$ are present,
    and a derivation $\aderiv'$ in $\aAPHS$ has been reached. 
    This nondeterministic procedure is strongly normalising.
\end{enumerate}

Apart from the notions of rule derivability and admissibility
defined above that
are based on the standard notion of consequence relation, 
in \cite{grab:2003} two variant notions are studied that refer to
variant consequence relations.
For these variant notions similar results are obtained;
their interconnections with each other and with the standard notions
are studied with the result of `interrelation prisms'.

\section{Abstract natural-deduction systems}
  \label{sec:ANDSs}

In this section the notion of rule in an `abstract natural-deduction system'
(\ANDS) is only hinted at. For the details I refer to Chapter~4 and Appendix~B
of my thesis \cite{grab:2005}. 

Rules in natural-deduction style proof systems typically have a more
complex form than rules in Hilbert systems:
\begin{Figure}[t]
  \begin{center}
    \scalebox{0.55}{\begin{picture}(0,0)%
\includegraphics{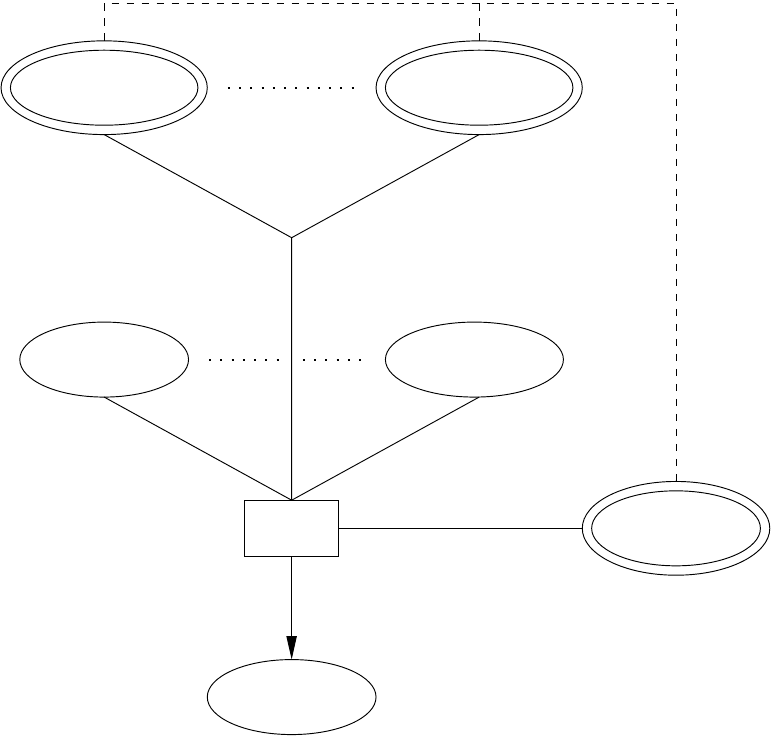}%
\end{picture}%
\setlength{\unitlength}{3947sp}%
\begingroup\makeatletter\ifx\SetFigFontNFSS\undefined%
\gdef\SetFigFontNFSS#1#2#3#4#5{%
  \reset@font\fontsize{#1}{#2pt}%
  \fontfamily{#3}\fontseries{#4}\fontshape{#5}%
  \selectfont}%
\fi\endgroup%
\begin{picture}(6166,5869)(2618,-5393)
\put(4276,-3061){\makebox(0,0)[b]{\smash{{\SetFigFontNFSS{12}{14.4}{\rmdefault}{\mddefault}{\updefault}{\color[rgb]{0,0,0}$1$}%
}}}}
\put(5626,-3061){\makebox(0,0)[b]{\smash{{\SetFigFontNFSS{12}{14.4}{\rmdefault}{\mddefault}{\updefault}{\color[rgb]{0,0,0}$n$}%
}}}}
\put(4276,-961){\makebox(0,0)[b]{\smash{{\SetFigFontNFSS{12}{14.4}{\rmdefault}{\mddefault}{\updefault}{\color[rgb]{0,0,0}$1$}%
}}}}
\put(5626,-961){\makebox(0,0)[b]{\smash{{\SetFigFontNFSS{12}{14.4}{\rmdefault}{\mddefault}{\updefault}{\color[rgb]{0,0,0}$n$}%
}}}}
\put(3451,-286){\makebox(0,0)[b]{\smash{{\SetFigFontNFSS{14}{16.8}{\rmdefault}{\mddefault}{\updefault}{\color[rgb]{0,0,0}$\pmassmi{1}{\ainst}$}%
}}}}
\put(6451,-286){\makebox(0,0)[b]{\smash{{\SetFigFontNFSS{14}{16.8}{\rmdefault}{\mddefault}{\updefault}{\color[rgb]{0,0,0}$\pmassmi{n}{\ainst}$}%
}}}}
\put(3451,-2461){\makebox(0,0)[b]{\smash{{\SetFigFontNFSS{14}{16.8}{\rmdefault}{\mddefault}{\updefault}{\color[rgb]{0,0,0}$\premi{1}{\ainst}$}%
}}}}
\put(6451,-2461){\makebox(0,0)[b]{\smash{{\SetFigFontNFSS{14}{16.8}{\rmdefault}{\mddefault}{\updefault}{\color[rgb]{0,0,0}$\premi{n}{\ainst}$}%
}}}}
\put(8026,-3811){\makebox(0,0)[b]{\smash{{\SetFigFontNFSS{14}{16.8}{\rmdefault}{\mddefault}{\updefault}{\color[rgb]{0,0,0}$\dmassm{\ainst}$}%
}}}}
\put(4951,-5161){\makebox(0,0)[b]{\smash{{\SetFigFontNFSS{14}{16.8}{\rmdefault}{\mddefault}{\updefault}{\color[rgb]{0,0,0}$\concl{\ainst}$}%
}}}}
\put(4951,-3811){\makebox(0,0)[b]{\smash{{\SetFigFontNFSS{17}{20.4}{\rmdefault}{\mddefault}{\updefault}{\color[rgb]{0,0,0}$\ainst$}%
}}}}
\end{picture}%
}
  \end{center}
  \caption{\label{fig:inst:ANDS-Rule}%
           Visualisation as an hypergraph hyperedge of
           an instance $\ainst$ with arity $\arity{\ainst} = n$ 
           of an \ANDS-rule.}
\end{Figure}
An instance $\ainst$ of a rule in a natural-deduction system 
is usually not exclusively determined by
a sequence of premises and a conclusion, but its description
frequently also involves, per premise,
a set of assumptions that has to be present, and a number of
assumptions that may be, or in fact are, discharged at $\ainst$.
Instead of assumption formulas, assumptions in natural-deduction systems
are often formalised as formulas with decorating markers 
that are used to single out those marked assumptions which are discharged 
at an instance.

Figure~\ref{fig:inst:ANDS-Rule} contains an illustration of
an instance of an unnamed rule for an `abstract natural-deduction system'
as defined in \cite{grab:2005}:
apart from premise and conclusion functions, unnamed \ANDS-rules
also contain functions $\spmassm$ and $\sdmassm$ that map instances
to their sequences of \emph{present marked assumptions} per premise,
and to their \emph{discharged marked assumptions}, respectively.

In a sequent-style representation, a typical instance $\ainst$ 
of an \ANDS-rule, one with $\arity{\ainst} = n$
as depicted in Figure~\ref{fig:inst:ANDS-Rule},
can be written in the following form:
\begin{equation}
  \label{eq:ANDS-rule:seqstyle}
  \mbox{
   \AxiomC{$ \pmassmi{1}{\ainst} \Rightarrow \premi{1}{\ainst} $}
   \AxiomC{\ldots}
   \AxiomC{$ \pmassmi{n}{\ainst} \Rightarrow \premi{n}{\ainst} $}
   \TrinaryInfC{$
     \bigl( \bigcup_{i=1}^n \pmassmi{i}{\ainst} \bigr) 
       \setminus \dmassm{\ainst}
         \,\Rightarrow\, \concl{\ainst}
                $}
   \DisplayProof
        }
\end{equation}
For the precise formal
definitions of rule derivability and admissibility in \ANDS{s}
some care is needed. The situation is considerably less straightforward 
than in \APHS{s}, and I refer to \cite{grab:2005} for the details.
I want to mention, however, that the definitions of admissibility and
derivability of \ANDS-rules in \ANDS{s} can be obtained in the 
following way:
(i) by considering sequent-style representations of \ANDS-rules
    with instances of the form~\eqref{eq:ANDS-rule:seqstyle} as \APHS-rules
    on sequents as formulas,
    and by considering sequent-style representations of entire \ANDS{s}
    as \APHS{s} with sequents as their formulas,
(ii) by applying 
     the definitions of rule derivability and admissibility in \APHS{s}
     to these \APHS-rules and \APHS{s},
and 
(iii) by transferring the resulting conditions back to \ANDS{s}. 
That is to say, rule derivability and admissibility in \ANDS{s}
can be defined by applying the definitions
of rule derivability and admissibility  in \APHS{s} 
for \APHS-rules on sequents that represent \ANDS-rules
with respect to \APHS{s} (with sequents as formulas)
that represent \ANDS{s}.

In my thesis, the definition of rule derivability and admissibility
for natural-deduction systems proved to be useful in the manner
indicated in the Section~\ref{sec:intro}.
In essentially all relevant cases of rules $\arule$ of the
Amadio--Cardelli system \ACeq\ whose status
in \BHeq\ earlier seemed doubtful to me,
I succeeded in proving that $\arule$ is admissible, but not derivable
in the Brandt--Henglein system~\BHeq; and consequently, that
instances of $\arule$ cannot just be simulated in \BHeq\ 
by mimicking derivations, but have to eliminated in an other,
typically more complicated, way.
Having earlier obtained  elimination procedures for instances
of such rules $R$ from \BHeq-derivations,
I at least obtained some certainty that I had not overlooked an
obvious way to mimic $\arule$-instances by derivations in \BHeq.

\section{Summary, a research idea, and thanks}
  \label{sec:conclusion}

\paragraph{Summary.}
By work for my Ph.D.~thesis on proof-theoretic interpretations into each other 
of proof systems for recursive types I became aware of the general relevance
of rule derivability and admissibility for interpretational proof theory. 
Due to the fact that rule derivability and admissibility are usually only
defined for concrete proof systems, and since the definitions of these
notions in natural-deduction style systems had not been clear to me
from the outset, I started a study of general properties of these notions
in abstractly viewed proof systems (both Hilbert\rule{1pt}{0pt} 
and natural-deduction style).
Formal definitions of abstract notions of inference rule
such as extensional rule descriptions, on which I had based
this study in my early attempts, 
were criticised by Roel as inadequate and, most of all, as conceptually 
unsatisfiable. 
Stimulated by Roel's remarks and his suggestion to try to draw inspiration from
the concept `abstract rewriting system' in rewriting theory,
I formulated the frameworks of \APHS{s} and \ANDS{s} for
abstract Hilbert-style and natural-deduction style proof systems.
In these systems, rules  
are treated as sets of (abstract) inference steps rather than
as mere relations between formulas, thereby keeping 
some information about how concrete rules are defined intensionally. 
In the sequel I used the concepts of \APHS{s} and \ANDS{s} to
state general results about derivability and admissibility of rules 
in pure Hilbert systems, and in natural-deduction systems.

\paragraph{A research idea.}
The frameworks of \APHS{s} and \ANDS{s} are based on abstract concepts of rules
that do not make use (at least not a priori) of special assumptions
on the formula language. 
While studying rule derivability and admissibility in these kinds
of abstract proof systems was sufficient for my particular purposes,
I think that it would also be fruitful to carry out a similar investigation
in abstract proof systems that are based on syntactic rule concepts,
perhaps 
formalised within a `logical framework' such as LF.
In particular, it would be interesting to see what kind of additional results 
about derivability and admissibility become possible 
once the syntactic manners in which inference rules are usually
defined is exploited systematically.
In the setting of proof systems that are formalised within a logical framework,
an idea for research would be to view elimination procedures for inference rules 
as higher-order rewriting systems, and to try to take advantage of 
the well-developed theory for higher-order rewriting. 

\paragraph{Thanks.}
I want to conclude by expressing my gratitude to Roel 
for his critical remarks about my earlier use of extensional rule descriptions,
his reference to abstract rewriting systems,
and the help he gave to me in regular discussions about the notions
of abstract proof systems during my last year as an \emph{AIO}
at the Vrije Universiteit, Amsterdam.

\paragraph{Acknowledgement.}
  I thank Vincent van Oostrom for discussions,
  for explaining to me the origins of abstract rewriting systems
  in the work of Newman,
  and for his comments about previous versions of this text.
  My thanks go also to Jan Willem Klop for indicating a number of
  typos and inaccuracies in language use.

\bibliography{main}

\end{document}